\documentclass[9pt,twocolumn,twoside]{opticajnl}
\journal{opticajournal} % use for journal or Optica Open submissions
\usepackage{pgf}
\usepackage{framed}
\usepackage{setspace}
\usepackage[utf8]{inputenc}
\DeclareUnicodeCharacter{03A3}{\Sigma}

\setboolean{shortarticle}{true}
\usepackage{lineno}
\title{Super resolving binary-source hypothesis testing with a double-clad fiber coupler}

\author[1, *]{John S. Wallis}
\author[1, 2]{David R. Gozzard}
\author[1]{Alex M. Frost}
\author[1,2]{Benjamin P. Dix-Matthews}
\author[1]{Nicolas Maron}
\author[1]{Joshua J. Collier}
\affil[1]{International Centre for Radio Astronomy Research, The University of Western Australia, Crawley WA 6009, Australia}
\affil[2]{ARC Centre of Excellence for Engineered Quantum Systems (EQUS), Department of Physics, The University of Western Australia, Crawley WA 6009, Australia}

\affil[*]{john.wallis@research.uwa.edu.au}

\begin{abstract}
We present a technique for binary spatial mode demultiplexing, using a double-clad fiber coupler as an optical mode sorter, for hypothesis testing for one or two point sources in an incident optical field. By directly coupling an optical field through a double-clad fiber coupler, we demultiplex the field into the fundamental mode and a superposition of higher-order modes. We use the ratio of multi-mode to single-mode power to distinguish between single and double point sources. In a tabletop demonstration of the technique, we demonstrate the capability to accurately identify the presence of two sources separated below the Rayleigh limit for relative brightnesses from 0~dB to $-$20~dB. For sources with less than 5~dB difference in their relative powers, our imaging protocol can correctly determine the presence of a second optical source even when the two sources have separations 50$\times$ smaller than the Rayleigh limit. These results highlight the potential of this technique as a simple tool for super-resolving classification of a pair of point emitters, especially in the context of astronomical imaging for binary systems.

\end{abstract}

\setboolean{displaycopyright}{false} % Do not include copyright or licensing information in submission.

\begin{document}

\maketitle

% \section{Introduction}

Classically, the resolution of an optical system has been heuristically characterized using Rayleigh's criterion \cite{LRayleigh:1879}. 
This criterion has provided a scaling metric for the performance of so-called "direct imaging" based on the imaging wavelength and the physical size of the optics used in the imaging system. When considering imaging in an astronomy context, science goals already dictate imaging wavelengths, and cost and engineering complexity typically limit the size of a system's primary optic. 

There exist imaging techniques with performance that scales better than the Rayleigh limit \cite{SHell:1994, HDefienne:2024}. However, many of these techniques require active illumination control to engineer specific states of light to improve the imaging performance. There have been several proposals for new types of imaging systems that use novel photonic techniques to image with resolution beyond that of the Rayleigh limit \cite{MTsang:2016, Paur:16, MPearce:2017, MTsang:2017, JRehacek:2017, AShan:2017, JRehacek:2018} in a passive context and, importantly for astronomy, using incoherent sources. Tsang \cite{MTsang:2016} proposed a multimodal imaging technique named "spatial mode demultiplexing" (SPADE), which could be used to achieve super-resolution by measuring the field in a quantum optimal spatial basis. For SPADE, a mode converter is required to measure the overlap of the incoming field with these basis modes; this has been achieved with Multi-plane light converters \cite{PBoucher:2020, LSantamaria:2023,XTan:2023, CRouviere:2024,  LSantamaria:2024, JWallis:2025}, Spatial light modulators \cite{JFrank:2023,pushkina2021superresolution}, and interferometric techniques \cite{FYang:2016, Gosalia_2023}, among others \cite{Darji:24}.

These techniques require sophisticated optical setups and many detectors, but in Tsang's original proposal \cite{MTsang:2016}, he noted that a simpler technique could be used for source classification. For source classification, most information is held in the difference between the fundamental mode and higher-order modes. As such, in a technique named Binary-SPADE, a comparison can be made just from the fundamental mode to the multi-mode field for quantum optimal source classification \cite{Lu_2018, ZHuang:2021,GMichael:2022, JZhang:2024,Schlichtholz:24}.  We present a novel and simple method for implementing this technique: the use of a double-clad fiber coupler (DCF-C) to decompose the incoming field into its fundamental Gaussian mode and a multi-mode mixture of the higher-order modes. By monitoring the powers in both the single and multi-mode outputs, we estimate whether a field contains one or two point sources. We provide a proof of principle of the technique in a tabletop demonstration. We show super-resolving performance when measuring incoherent sources with spacings below the equivalent Rayleigh limit of our system. To our knowledge, this is the first experimental demonstration of a Binary-SPADE system \textcolor{black}{using the originally proposed protocol in \cite{MTsang:2016}, which decomposes the field into the fundamental Gaussian and an incoherent combination of all the higher order modes}.

% \section{Methods}

% \subsection{Experimental Setup}

\begin{figure*}
    \begin{center}
        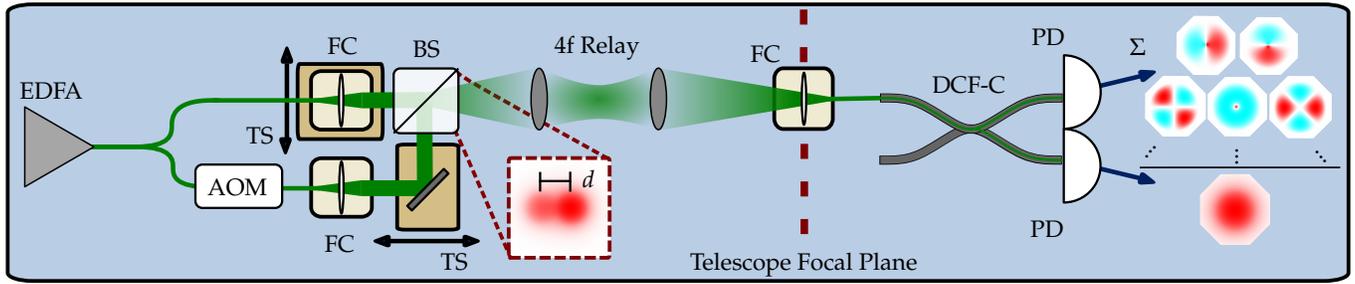
        \caption{Experimental Setup for source preparation and Binary SPADE-based source classification.  EDFA, Erbium-doped fiber amplifier; AOM, acousto-optic modulator; FC, fiber-collimator; TS, Translation stage; BS, beam splitter; DCF-C, Double-clad fiber coupler; PD, photodetector.}
        \label{fig:ExpSetup}
    \end{center}
\end{figure*}

% \textcolor{black}{Note --- Ooptics letters don't have sections headings. You still write it as though you have these sections, but you just get rid of the headings.}

Figure \ref{fig:ExpSetup} shows a schematic representation of the optical setup used for our experiment. Broadband spontaneous emission of an erbium-doped fiber amplifier (EDFA) operated without a seed laser is used to generate pseudo-thermal light. To characterize the coherence of this light, we measured the autocorrelation with a Mach-Zehnder interferometer. We found that interference disappeared with less than 500~\textmu m  path length difference. The path length difference of the two arms in our imaging experiment is orders of magnitude greater than the coherence length, so there should be no effect on the system due to mutual coherence between the sources.

After generation, the light from the source EDFA is split with a fiberized beam splitter. One path is directly collimated into free space, while the second path is first sent through an Acousto-optic modulator (AOM) used to provide variable attenuation before it is collimated into free space. Both arms impinge on a beam splitter using mirrors mounted on motorized translation stages. By moving the mirrors, the beams can be overlapped or separated by some displacement $d$.

Light from the beam splitter is imaged onto a fiber collimator using a 4f relay. This fiber collimator is used to couple light into a DCF routed to a DCF-C (Thorlabs DC1300LE2FA), which couples the guided fiber modes into single and multi-mode fibers. Using two photodiodes, we can monitor the power in both the fundamental and higher-order modes, which we use for hypothesis testing. To align the DCF to the imaging system, we tune the alignment of each beam for the maximum ratio between the single-mode port to the multi-mode port (isolation). Isolation between ports is a limiting factor in the imaging system's performance, and by optimizing for maximal isolation, we optimize the performance of our protocol. \textcolor{black}{The maximum achieved isolation in our experiment is 10.5 dB.}

% \subsection{Stage Calibration}

To verify the capabilities of this setup to accurately provide a source configuration with a given displacement $d$ and power ratio $\epsilon$, we verified that the optical loss of the AOM varies by 1.000 dB/dB of RF power over the range of powers used in our experiment. We also performed a calibration to ensure that the motorized translation stages accurately and repeatably produce the necessary displacements. To check this, we compare the positioning of the stage to the phase readout of an external Michelson interferometer with one arm reflecting from an additional mirror mounted to the translation stage. Comparing these measurements and the repeatability in the response of the measured DCF-C, our results are consistent with the stages' specified repeatability of $<$5 $\mu m$.

% \subsection{Imaging Protocol}

As a proof of principle, we use the double-clad fiber to determine if the imaged electric field was produced by one or two point sources. To simulate a realistic telescope system with a finite aperture, we use the output Gaussian beam waist $1/e^2$ radius of 1.5~mm as a proxy for the point spread function (PSF) that would limit the resolution of a realistic telescope system, \textcolor{black}{In an ideal system this would be an Airy Disk which we approximate with a Gaussian \cite{MTsang:2016}}. 
\textcolor{black}{We design our optical system, by selecting both the fibers and magnification in our relay,} so that the fundamental mode of the DCF matches the proxy PSF produced by our source. As such, a single source should couple almost completely into the single-mode port of our imaging system, while a source composed of multiple sources will couple partially into both the single-mode and multi-mode ports. For comparison, the James Webb Space telescope has a PSF FWHM on the order of 10~\textmu m at a wavelength of $1550$~nm \cite{JWSTdocs}, which closely matches the single-mode mode field diameter of the DCF.

To perform a hypothesis test, we first calibrate the system with one beam blocked to determine the response to a single source. We integrate the photocurrent of the photodiodes over 10~s to measure optical power. We translate the beam with the motorized translation stages to a random position and then back to be centered on the optical axis. In doing so, we can estimate the variation in power we expect to see from one source due to alignment errors. We repeat this procedure 10 times and measure across a range of misalignment up to 0.1 waist radii (wr). We then use the worst-case ratio between multi-mode to single-mode coupling as our calibrated single-source baseline response. Our protocol then compares the multi-mode to single-mode ratio to this baseline, in order to perform a hypothesis test, which determines if an imaged field was produced by one or two point sources. \textcolor{black}{We assume any measurement we do not classify as a single point source to be a binary.} \textcolor{black}{Using 1~s integrations, we found a 1.6~\% chance to misclassify a single source as a binary, but found errors only occurred when misaligned by 0.04~wr. This error probability dropped to 0~\% when integrating for longer than 10~s}. By calibrating this way, we ensure we minimize the chance of a single source being classified as two sources, even in the presence of pointing errors typical of modern telescope systems (up to 10 \% \cite{JWSTdocs} from optical center of intensity).

To test the performance of our protocol, we measure the response of the DCF-C when the two sources are separated across a range of separations from 20~\textmu m up to 6~mm (0.013 to 4 PSF radii) and with the ratio of the source powers ranging from 0~dB to 20~dB. When positioning the beams for a source configuration, we translate each beam relative to their power to ensure the optical center of mass remains aligned to the imaging system's optical axis. In a practical system, this pointing method is easily achievable via direct imaging and is already how most telescope systems align their pointing. For each source configuration, we integrate the photocurrent over 10~s to measure the optical power. We repeat the measurement of each source configuration 10 times, resetting the experiment by zeroing the source position and power ratios between measurement runs.

% \section{Results}

\begin{figure*}
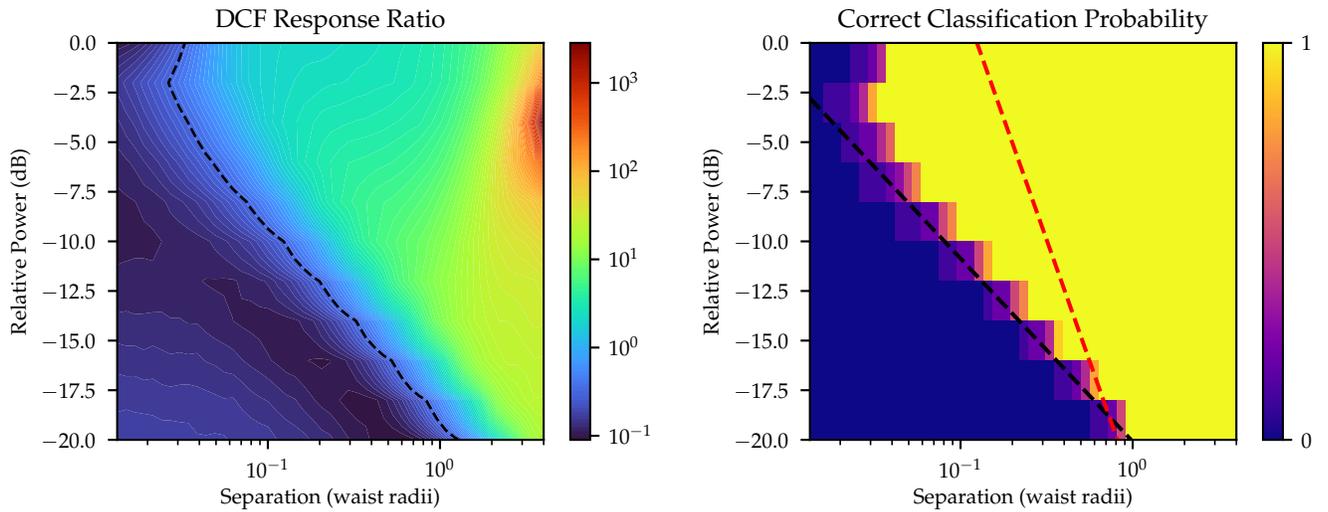

    \begin{center}
     \input{ResponseContoursWithThreshold.pgf}
    \input{ClassificationWithDI.pgf}

    \caption{Left: Contour plot of the ratio of the coupled power in the multi-mode port to the single-mode port as a function of the relative power of the two sources (y-axis) and separation between them (x-axis). As expected, larger separation results in a larger multi-mode to single-mode coupling ratio for a given power ratio. \textcolor{black}{The black dashed line shows the threshold we calibrated to classify a source as a binary.} Right: Hypothesis test of whether the imaged source was created by one or two point sources. The color indicates how likely our system is to classify the source correctly. The black dashed curve is fit to the transition from correct to incorrect classification. \textcolor{black}{The red dashed curve is the simulated performance limit of ideal direct imaging.}}
    \label{fig:ResponseContours}
    \end{center}
\end{figure*}

Figure \ref{fig:ResponseContours} (Left) shows the response ratio of the DCF imaging system as a function of relative power ratio (y-axis) and separation (x-axis).  We compare the double source response ratio to the calibrated single source response ratio to perform a hypothesis test for if the field was generated using one source or two sources. Figure~\ref{fig:ResponseContours} (Right) shows the probability of a correct classification as a function of relative power ratio (y-axis) and separation (x-axis). 

% \begin{figure}
%     \begin{center}
%      \input{Classification.pgf}
%     \caption{Hypothesis test of whether the imaged source was created by one or two point sources. The color indicates how likely our system is to correctly classify the source. The black dashed curve is fit to the transition from correct to incorrect classification.}
%     \label{fig:HypothesisTest}
%     \end{center}
% \end{figure}

\begin{figure}
\input{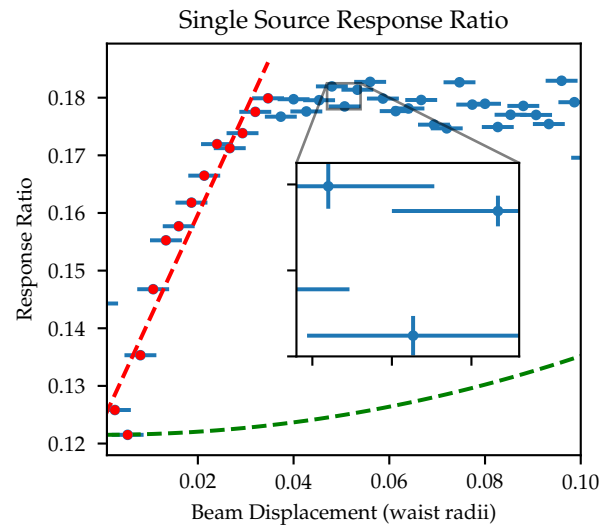}
        \caption{\textcolor{black}{Multi-mode to single-mode} response ratio of the DCF when a single source is translated a small amount from the center of the imaging system. Note that vertical error bars are too small to show on the plot. \textcolor{black}{The red dashed line is a linear fit to the data below 0.04~wr. The green dashed line is the expected coupling ratio given crosstalk.}}
        \label{fig:SmallSignal}
    
\end{figure}

To further evaluate the performance of our system, we characterized the sensitivity of our system to small displacements by translating a single source across a range of 150~\textmu m (0.1 wr). This allows us to determine how the pointing accuracy of a telescope will limit the performance of a Binary-SPADE system. We present these results in Figure~\ref{fig:SmallSignal}. 

% \section{Discussion}

In Figure \ref{fig:ResponseContours} (Left), we can see the direct response of the measured mode ratio in our experiment. As expected, we see generally low coupling in the multi-mode port relative to the single-mode port for small separations. As separation (x-axis) increases, the Gaussian extent of the sources does not overlap with the mode of the single-mode fiber, and thus the multi-mode to single-mode coupling ratio increases.  When imaging sources with differing powers, our imaging protocol will center the fundamental fiber mode on the optical center of mass. Due to the centering procedure, when we compare sources with the same separation, but with a greater difference in relative power, the brighter source will have more overlap with the imaging system's fundamental mode.  We can see this effect in the results -- the multi-mode to single-mode coupling ratio decreases as the difference in brightness of the two sources increases. The interaction of these two effects leads to the diagonal regions of constant response ratio seen in the plot. Notably, the peak coupling ratio occurs with a slight power mismatch between the sources; this is due to residual misalignment between the source and the center of the DCF in our experiment. 

Due to the degenerate response of this imaging system, it is not possible to simultaneously estimate the relative power of the sources and the separation. However, the DCF can still be used for hypothesis testing if the measured optical field is produced by one or two sources. Figure~\ref{fig:ResponseContours} (Right) shows the result of using these measured response ratios to perform such a test. For an equivalent direct imaging setup, a waist separation of 1.22~wr corresponds to the Rayleigh Limit. From this plot, it can be seen that the DCF is capable of super-resolving some source configurations with a power ratio between the dimmest to brightest source as low as -20 dB. Due to the effect of pointing at the optical center of mass, the response ratio does not change as significantly for increasing separations for sources with larger power ratios; as such, this degrades the performance of the classification protocol. Fitting to the transition from accurate to inaccurate classification in Figure~\ref{fig:ResponseContours} (Right), (i.e., the edge of the yellow region), the smallest discernible power ratio decreases by 9.23~dB/decade of separation between the sources.
At relative powers (i.e., more equal brightnesses) with less than a 5~dB difference, the system has a significant increase in performance compared to the diffraction limit. Figure~\ref{fig:ResponseContours} (Right) shows error-free classification for separations more than 50$\times$ smaller than the Rayleigh limit. 

From Figure~\ref{fig:SmallSignal}, \textcolor{black}{while one would expect a smooth change in coupling as the fiber and source modes are misaligned, we see an initial sharp change in coupling. \textcolor{black}{The knee point at 0.04~wr corresponds to the minimum misalignment to misclassify a single source.} We postulate this effect to be due to tolerance in both the DCF and DCF-C manufacturing, which results in non-ideal coupling.} We use this response to inform how pointing tolerance affects the classification response of this system.  A better-pointing tolerance results in typically lower response ratios for single sources. This allows a lower threshold to be used for classifying a double source configuration. As such, a better-pointing tolerance should result in sources with lower separation and a greater difference in their powers to be successfully classified. As a point of comparison, the James Webb Space Telescope can achieve a pointing accuracy relative to its PSF of 9.15\% \cite{JWSTdocs}. We have used the worst-case scenario measured with 10\% misalignment, so the pointing accuracy of contemporary telescopes will not degrade our technique. The performance of the Binary-SPADE system presented in this Letter is primarily limited by the isolation between the single-mode and multi-mode fiber. DCF-Cs with better separation and isolation of the fundamental mode from the higher-order modes will reduce the noise in the multi-mode-to-single-mode ratio measurement, improving the accuracy of the source discrimination. \textcolor{black}{Throughout this analysis and the experiment, we have assumed the PSF is a Gaussian, to approximate an ideal diffraction-limited PSF. In realistic systems, imaging aberrations will result in a PSF that need not resemble a Gaussian, or even be rotationally symmetric. This is true in the case of many real telescope systems, including the James Webb Telescope \cite{JWSTdocs}. In such a case, a point source will no longer couple exclusively to the fundamental mode of the DCF and will result in some unavoidable crosstalk between the single and multi-mode ports. This crosstalk will result in some degradation of the performance \cite{MGessener:2020}. Further investigation into the typical degradation due to this effect will be necessary before applying this technique in realistic telescope systems.}

% \section{Conclusion}

We have provided an experimental demonstration of Binary-SPADE, a simple technique for classifying whether an image contains one or two sources. Our results show that by using a DCF-C to perform a simple mode decomposition, super-resolving classification of the presence of one or two sources can be achieved. While this work has only demonstrated the case of one versus two sources, this technique should apply to discriminating between a single object and the presence of any extended source, such as multi-object fields, rings, or irregular and non-symmetrical distributions. Immediate follow-up studies from this work should aim to determine how to implement this technique to classify images composed of an arbitrary number of sources or estimate some of the parameters, i.e., relative power or separation of the sources \textcolor{black}{and test sources in the photon-counting regime}. Due to the simplicity of the technique, this could serve as a useful platform for future proof-of-principle demonstrations of passive super-resolution techniques, necessary to advance the field, including in realistic on-sky astronomical imaging experiments. \textcolor{black}{Relative to other SPADE techniques, this has very low photonic loss with losses dominated by fiber coupling (typically around 1~dB loss) and the loss due to the DCF-C (<0.7 dB in our experiment).} If the performance of Binary-SPADE achieved in this work can be replicated on-sky, the integration of a Binary-SPADE system onto existing and future large telescopes would have useful scientific applications in the identification and study of binary systems and the measurement of the angular size of stars or other objects \cite{ZHuang:2021}. The recent discovery that a known black hole X-ray binary system is a trinary system \cite{burdge2024black}, which relied on accurately measuring the presence of the two luminous sources in the system, demonstrates how a telescope employing Binary-SPADE could make immediate contributions to the study of the evolution of black hole systems. Practically, in purpose-built super-resolving telescopes, imaging protocols that can utilize more modes will likely provide performance advantages over Binary-SPADE, namely their demonstrated ability to provide accurate parameter estimation \cite{VAnsari:2021} and image sources of greater spatial complexity \cite{pushkina2021superresolution}.

% \section{Back matter}

\begin{backmatter}
\bmsection{Funding} Australian Research Council (project ID CE17010009, project ID DE240100587), Air Force Office of Scientific Research (project ID FA2386-23-1-4081).

\bmsection{Acknowledgment} J.S.W, A.M.F \& J.J.C are supported by Australian Government Research Training Program Scholarships. This material is based upon work supported by the Air Force Office of Scientific Research under award number FA2386-23-1-4081. The Authors would also like to thank Dr. Andrew Lance, Dr. Shane Walsh, and Dr. Elrina Hartman for helpful conversations.

\bmsection{Disclosures} The authors declare no conflicts of interest.

\bmsection{Data Availability Statement} Data underlying the results presented in this paper are not publicly available at this time but may be obtained from the corresponding author upon reasonable request.

\end{backmatter}

\bibliography{references}
% Full bibliography added automatically for Optics Letters submissions; the following line will simply be ignored if submitting to other journals.
% Note that this extra page will not count against page length
\bibliographyfullrefs{references}

\end{document}